\documentclass[a4,aps,amsmath]{revtex4}
\usepackage{graphicx}
\usepackage{color}
\newcommand{\be}{\begin{equation}}
\newcommand{\ee}{\end{equation}}
\newcommand{\ba}{\begin{array}}
\newcommand{\ea}{\end{array}}
\newcommand{\bea}{\begin{eqnarray}}
\newcommand{\eea}{\end{eqnarray}}
\newcommand{\bdm}{\begin{displaymath}}
\newcommand{\edm}{\end{displaymath}}

\begin{document}

\title{Hysteresis in Random Field XY and Heisenberg Models: \\ Mean 
Field Theory and Simulations at Zero Temperature}

\author{Prabodh Shukla}
\email{shukla@nehu.ac.in}
\affiliation{%
Physics Department \\ North Eastern Hill University \\ 
Shillong-793 022, India}%
\author{R S Kharwanlang}
\affiliation{%
Physics Department \\ North Eastern Hill University \\ 
Shillong-793 022, India}%


\begin{abstract}

We examine zero temperature hysteresis in random field XY and 
Heisenberg models in the zero frequency limit of a cyclic 
driving field. Exact expressions for hysteresis loops are 
obtained in the mean field approximation. These show rather 
unusual features. We also perform simulations of the two 
models on a simple cubic lattice and compare them with the 
predictions of the mean field theory.

\end{abstract}

\maketitle

\section{Introduction}

Random field XY and Heisenberg models provide a simple 
framework for exploring the effects of quenched disorder in 
classical systems of continuous symmetry 
~\cite{imry-ma,young}. These models and their variants have 
helped in understanding a wide range of phenomena including 
random pinning of spin and charge density waves in metals
~\cite{biljakovic,strogatz1,gruner}, vortex lattices in 
~disordered
type-II superconductors ~\cite{fisher,giamarchi}, liquid 
crystals in porous media 
~\cite{ianna,maritan,crawford,buscaglia}, and disordered 
ferromagnets 
~\cite{silveira1,silveira2,dieny,pierce1,pierce2,jagla}. 
Initially the models were used to understand the effect of 
disorder on equilibrium properties of materials. However, 
with shifting trends in statistical mechanics towards 
non-equilibrium phenomena, the same models have been 
supplemented with a simple relaxation dynamics and adapted to 
study non-equilibrium behavior of systems including their 
response to a driving field. In the present paper we examine 
zero temperature hysteresis in these models when the 
frequency of the cyclic driving field goes to zero, and 
provide an exact solution of the hysteresis loop in the mean 
field limit. The zero temperature dynamics is deterministic 
and therefore simpler to analyze theoretically. But this is 
not the only reason for using it. It is also meaningful for 
describing disorder-driven hysteresis in real materials at a 
finite temperature. Materials with quenched disorder are 
characterized by a large number of metastable states 
separated from each other by energy barriers that are much 
larger than the thermal energy of the system. These findings 
are based on extensive studies of spinglasses and other 
random field systems
~\cite{young} but are intuitive as well. Intuition tells us 
~that if the
disorder remains frozen over experimental time scales, 
thermal energy must be smaller than the barriers due to 
disorder. Similarly metastable states must be aplenty because 
most of these have an apparently random configuration. A 
random configuration does not bring to mind any specific 
configuration but rather a large number of possible 
configurations.

Zero temperature hysteresis in XY and Heisenberg models in 
the zero frequency limit of driving field has been studied by 
Rava da Silveira and Mehran Kardar ~\cite{silveira1} as well. 
They use a slightly different variant of the model than the 
one studied here. The random field in their model has a 
Gaussian distribution centered at zero. In our model, the 
random fields are in the form of randomly oriented unit 
vectors. We determine the hysteretic response of the system 
to a changing field by solving the equations of motion 
directly for a given initial condition. Silveira and Kardar 
take an indirect approach. They recast the equations of 
motion into a path integral. The path integral is a sum over 
all paths of the exponential of an action. It includes paths 
corresponding to different initial conditions. Silveira and 
Kardar employ a method for extracting the physically relevant 
hysteretic path from amongst multiple solutions. We refer the 
reader to reference ~\cite{silveira1} for details. The main 
object of their study is to examine critical points in the 
hysteretic response of a system. They focus on a point on the 
hysteresis loop where the susceptibility of the system 
diverges. If there is such a point on one half of the 
hysteresis loop, say in increasing applied field, there is 
also a symmetrically placed point on the other half of the 
loop corresponding to decreasing field. These points are 
called non-equilibrium critical points because they are 
characterized by a diverging correlation length, and show 
scaling of various quantities and universality of critical 
exponents that is reminiscent of equilibrium critical point 
phenomena. Sethna et al ~\cite{sethna1,sethna2,dhar} studied 
the non-equilibrium critical points on the hysteresis loop in 
the random field Ising model with a Gaussian distribution of 
random fields. Silveira and Kardar ~\cite{silveira1} ask the 
question if the universality class of critical hysteresis 
studied by Sethna et al would change if we go from Ising 
spins to vector spins. They find no change in the case when 
the critical point occurs at a nonzero value of either the 
applied field or the magnetization. However, if the critical 
point were to occur when both the applied field and the 
magnetization vanish, all components of the order parameter 
may become critical simultaneously. In this case the critical 
point would have full rotational symmetry of vector spins 
with a new set of critical exponents.

We focus on the shape of hysteresis loop rather than the critical points 
on it. The shape of hysteresis is not a universal object like a set of 
critical exponents, but nonetheless it is of practical interest. An 
exact calculation of hysteresis loop also determines if there are first 
order jumps or critical points on the loop. The calculations presented 
here bring out two rather unexpected but interesting results. In the 
random field XY model, there is a window in the value of the 
ferromagnetic coupling parameter where the hysteresis loop splits into 
two small loops at large values of the cyclic field but there is no 
hysteresis at small values of the field. This prediction of the mean 
field theory is also seen qualitatively in our simulations of the model 
on simple cubic lattices with nearest neighbor interactions. Similar 
shapes have been observed earlier in the random field 
Blume-Emery-Griffiths model for martensitic transitions 
~\cite{goicoechea}, and other theoretical models and experiments 
\cite{dieny,pierce1,pierce2,jagla} but they do not appear to be known 
very widely. The other point is that our mean field theory predicts a 
different kind of phase transition in random field XY model than in 
random field Heisenberg model. This is somewhat surprising at first 
sight because the two models have the same critical behavior in the mean 
field limit of the renormalization group theory. However, it is 
understandable if we keep in mind that our model has a different 
distribution of the random field than the one used in reference 
~\cite{silveira1}. We shall return to this issue after presenting our 
results.

The present paper is organized as follows. In section II, we 
explain the model based on $n$-component unit vector spins, 
and zero temperature dynamics. A discrete-time equation of 
motion for the magnetization along the applied field is also 
set up in this section. Section III is devoted to the 
analysis of the shapes of hysteresis curves. It has four 
sub-sections. Sub-sections A and B are devoted to the 
determination of hysteresis loops in the mean field theory 
for XY and Heisenberg models respectively. Sub-section C 
takes a closer look at the nature of criticality in the mean 
field theory of hysteresis. Sub-section D presents the 
results of numerical simulations of the model on simple cubic 
lattices and comparison of these results with the predictions 
of the mean field theory. Section IV contains some concluding 
remarks.

\section{The model}

We consider the Hamiltonian

\be H=-J\sum_{i,j} \vec{S_i}.\vec{S_j} -\sum_{i} 
\vec{h_i}.\vec{S_i}-\vec{h}.\sum_{i}\vec{S_i} \ee

Here $\vec{S_{i}}$ and $\vec{h}_i$ are $n$-component unit 
vectors located at site-$i$ ($i=1,2,\ldots N$) of a 
d-dimensional lattice. In the context of magnetic systems, 
$\{\vec{S}_i\}$ are classical spins, $\{\vec{h}_i\}$ a set of 
on-site random fields, and $\vec{h}$ is a uniform applied 
field of magnitude $|h|$. We focus on $n=2$ (XY spins), and 
$n=3$ (Heisenberg spins) since the case $n=1$ (random field 
Ising model) has been studied rather thoroughly
~\cite{sethna2} albeit for a Gaussian distribution of the 
~random field. The summation over $j$ on the right-hand-side
(rhs) is restricted over the nearest neighbors of site-$i$. 
The first and the third terms on the rhs promote uniform 
order in the system: $J$ ($J>0$) is ferromagnetic exchange 
interaction that aligns nearest neighbors parallel to each 
other; the external field aligns each spin $\vec{S_i}$ along 
$\vec{h}$. The second term on the rhs disorders the system by 
attempting to align each spin $\vec{S_i}$ in a random 
direction $\vec{h_i}$. The random fields $\{\vec{h}_i; 
|\vec{h}_i|=1 \}$ are quenched, i.e. they do not evolve in 
time. The spins $\{\vec{S_i(t)}\}$ are the dynamical degrees 
of freedom.

At zero temperature, an initial configuration 
\{$\vec{S}_i(0)$\} evolves in time so as to lower the energy 
of the system. The evolution ends when each $\vec{S}_i(t)$ is 
aligned along the local field $\vec{f}_i(t)$ at that site. 
Let \{$\vec{S}_i^*$\} denote a configuration at the 
termination of the zero-temperature single-spin-flip 
dynamics. We call it a fixed point configuration because it 
remains unchanged under the dynamics. In the absence of 
disorder, the fixed point has all spins parallel to each 
other irrespective of the starting point \{$\vec{S}_i(0)$\}. 
This corresponds to the lowest energy of the system. In the 
presence of random fields $\{\vec{h_i}\}$, the fixed point 
becomes rather non trivial on two accounts. First, it may and 
generically does lose its translational symmetry. Second, it 
is no longer independent of the starting point. There is now 
a large set of fixed points each with its domain of 
attraction from where it can be reached. Each of these fixed 
points is a local minimum of energy. A local minimum is a 
stable state at zero temperature because there is no 
mechanism of escape from it unless the applied field is 
jacked up sufficiently. It would correspond to a metastable 
state under finite temperature dynamics if the thermal energy 
is smaller than the barriers of disorder, but we consider 
zero temperature dynamics only. In equilibrium problems with 
quenched disorder, one needs to know the lowest of the local 
minima. This is a difficult task analytically or 
computationally. Fortunately, the problem of hysteresis does 
not require the knowledge of the global minimum. Hysteresis 
is determined by the sequence of local minima visited by the 
system as it tries to follow a changing field. Our object is 
to determine this sequence as the applied field is cycled 
from $-\infty$ to $+\infty$ and back to $-\infty$ in small 
steps. At each step, we allow the zero temperature dynamics 
as much time as it requires to come to a fixed point.

We obtain the local minima by using a discrete time dynamics 
that progressively lowers the energy of the system. The 
dynamics transforms a spin configuration $\{ \vec{S}_i(t)\}$ 
at time $t$ into a lower energy configuration $\{ 
\vec{S}_i(t+1)\}$ at time $t+1$. The fixed point of this 
iterative procedure corresponds to a local minimum of the 
energy of the system. Our dynamics can be stated in a simple 
form if we rewrite equation (1) in terms of a local effective 
field $\vec{f}_i(t)$ at site-$i$,

\be H= -\sum_{i}\vec{f_i}(t).\vec{S_i}(t);\mbox{\hspace{.5cm}
} \vec{f_i}(t)=J \sum_{j}\vec{S_j}(t)+h_i+h \ee

The dynamics is given by the equation:

\be \vec{S_i}(t+1) = \frac{\vec{f_i}(t)}{|\vec{f_i}(t)|} \ee

At each site, a new spin $\vec{S}_i(t+1)$ is obtained that 
points in the direction of the local field $\vec{f}_i(t)$ at 
that site. The denominator in equation (3) ensures that the 
new spin $\vec{S}_i(t+1)$ has unit length; $\vec{S}_i(t+1)$ 
is therefore a rotated form of $\vec{S}_i(t)$. The rotation 
lowers the energy of each spin, and therefore that of the 
entire system. However, after the spins are rotated the local 
field changes as well. Thus the rotated spin $\vec{S}_i(t+1)$ 
is generally not aligned along the new local field 
$\vec{f}_i(t+1)$ at site $i$. We can reduce the energy of the 
system further by repeating the dynamics. Indeed, we start 
with a random initial configuration \{$\vec{S}_i(0)$\} and 
subject it to repeated applications of equation (3) until an 
attractor of the dynamics is reached. In discrete time 
dynamics, an attractor could be in principle a fixed point or 
a limit cycle. However in our analysis as well as 
simulations, we find the dynamics always reaches a fixed 
point configuration \{$\vec{S}_i^*$\}. The fixed point 
configuration corresponds to a local minimum of energy. The 
initial configuration \{$\vec{S}_i(0)$\} and the 
configurations along the path to the fixed point lie in the 
domain of attraction of the fixed point.

For simplicity, we characterize each configuration of spins 
by a single parameter that measures the magnetization of the 
system along the applied field $\vec{h}$. We assume that the 
applied field $\vec{h}$ is along the $x$-axis. The equations 
of motion for the magnetization of Heisenberg and XY spins 
are quite similar. We first consider the case of XY spins. In 
this case, $\vec{S}_i(t)$ and $\vec{h}_i$ can be completely 
specified by the angles $\theta_i(t)$ and $\alpha_i(t)$ that 
they make with the $x$-axis. The $x$-component of equation 
(3) gives,

\be \left. \cos \theta_i(t+1) = \frac{J \sum_j \cos 
\theta_j(t)+ h + \cos \alpha_i}{[ (J \sum_j \cos \theta_j(t) 
+ h + \cos \alpha_i)^2 + (J \sum_j \sin \theta_j(t) + \sin 
\alpha_i)^2 ]^{\frac{1}{2}}} \right.  \ee

The above equation is rather difficult to solve analytically 
except in the mean field limit when a site-$i$ interacts with 
every other site-$j$ of the system ($j \ne i$) with strength 
$J=J_0/N$. Let $S_i^x$ and $S_i^y$ be the components of XY 
spin $\vec{S_i}$ along the x and y axes respectively. We look 
for a solution of equation (4) in the case when the spins may 
be ordered along the x-axis, but there is no global ordering 
in the system in the $y$ direction. We write,

\be J \sum_j S_j^x(t)=\frac{J_0}{N} \sum_j S_j^x(t) = J_0 
\cos \theta(t)=J_0 m(t);\hspace{.25cm} \sum_j S_j^y(t)=0. \ee

The above equation defines a time dependent order parameter 
$\cos \theta(t)$, or equivalently a magnetization $m(t)=\cos 
\theta(t)$ as the average value of the component of 
$\vec{S_i}(t)$ along the applied field $\vec{h}$. We shall 
mostly use the notation $m(t)$, but keep $\cos \theta(t)$ for 
occasional use when convenient to do so.

Substituting from equation (5) into equation (4) we get,

\be \left. \cos \theta_i(t+1) = \frac{\{J_0 m(t) + h\} + \cos 
\alpha_i}{[ 1+ 2 \{J_0 m(t)+ h\} \cos \alpha_i + \{J_0 m(t)+ 
h\}^{2}
]^{\frac{1}{2}}} \right.  \ee

Equation (6) has a nice geometrical interpretation suggested by Mirollo 
and Strogatz ~\cite{mirollo} who analyzed the fixed point equations for 
the XY model for $h=0$ rather than the time dependent equation for 
$m(t)$. Note that the quantity $J_0 m(t)+h$ is the mean field trying to 
align $\vec{S}_i(t)$ along the $x$-axis. The mean field has the same 
value at each site. The random field at each site has a component equal 
to $\cos \alpha_i$ that (depending upon the sign of $\cos \alpha_i$) 
supports or opposes the alignment of $\vec{S}_i(t)$ along the $x$-axis. 
A geometrical relationship between the angles $\theta_i(t)$, $\alpha_i$, 
and the mean field at time $t$ is illustrated by figure (1) which shows 
two unit vectors separated from each other by a distance $J_0 m(t)+h$ 
along the $x$-axis, and making angles $\theta_i(t+1)$ and $\alpha_i$ 
respectively with the $x$-axis. From the geometry of figure (1), we may 
write

\be \tan \theta_i(t+1) = \frac {\sin \alpha_i }{[J_0 
m(t)+h+\cos \alpha_i]} \ee

Also, a well known identity relating the sines of the angles 
of a triangle to the its sides gives,

\be \sin \theta_i(t+1) = \frac {\sin [ \alpha_i 
-\theta_i(t+1) ]}{[J_0 m(t)+h]} \ee

Equation (6) is the most convenient form for studying the 
evolution of the order parameter $m(t)$ but equations (7) and 
(8) are useful to get a geometrical picture of the spin 
configuration of the system. For example, in the limit $ J_0 
m(t)+h \rightarrow 0$, equation (8) gives $\theta_i 
=\alpha_i$ as may be expected. If $J_0 +m(t)=1$, equation (8) 
gives $\theta_i(t+1)=\alpha_i/2$. This is expected as well. 
In this case the mean field as well as the random field have 
unit magnitude. One acts along the $x$-axis and the other 
makes at an angle $\alpha_i$ with the $x$-axis. Therefore the 
resultant field aligns the spin at an angle $\alpha_i/2$ with 
the $x$-axis.

We obtain a recursion relation for $m(t+1)=\cos \theta(t+1)$ 
by averaging equation (6) over all sites,

\be m(t+1) =\left. \frac{1}{2\pi}\int_0^{2\pi} \frac{\{J_0 
m(t) + h\} + \cos \alpha_i}{[ 1+2 \{J_0 m(t)+h\} \cos 
\alpha_i +\{ J_0 m(t) +h\}^2
]^{\frac{1}{2}}} d \alpha_i \right. \mbox{ [XY model] } \ee

A similar mean field equation is obtained for the Heisenberg 
model. A Heisenberg spin $\vec{S}_i(t)$ may be specified by 
an azimuthal angle $\phi_i(t)$ that the spin makes from a 
fixed axis (say the $y$-axis) in the $yz$ plane and the polar 
angle $\theta_i(t)$ that it makes with the $x$-axis. The 
random field $\vec{h_i}$ is also to be specified by a polar 
angle $\alpha_i(t)$, and an azimuthal angle $\psi_i(t)$. As 
in the case of the XY model, we assume that the field 
$\vec{h}$ is applied in the $x$-direction, and any global 
order in the system lies along the $x$-direction only.

\be J\sum_j S_j^x(t)=\frac{J_0}{N} \sum_j S_j^x(t)= J_0 \cos 
\theta(t)=J_0 m(t);\hspace{.25cm} \sum_j S_j^y(t)=0; 
\hspace{.25cm}\sum_j S_j^z(t)=0, \ee

This gives us the following mean field equation for the 
Heisenberg model analogous to equation (9) for the XY model:

\be m(t+1) =\left. \frac{1}{4\pi}\int_0^{2\pi} d \psi_i 
\int_0^{\pi} \frac{\{J_0 m(t) + h\} + \cos \alpha_i}{[ 1+2 
\{J_0 m(t)+h\} \cos \alpha_i +\{ J_0 m(t) +h\}^2 
]^{\frac{1}{2}}} \sin \alpha_i d \alpha_i \right. \mbox{ 
[Heisenberg model] } \ee

\section{Hysteresis}

We use the dynamics described above to obtain magnetization 
curves in a slowly varying cyclic field. The field is 
increased from $h=-\infty$ to $h=\infty$ and then decreased 
to $h=-\infty$ so very slowly that the system has sufficient 
time to settle into a local minimum of energy at each point. 
In practice we start with a large negative field when the 
stable configuration of the system has all spins aligned 
along the negative $x$-axis, and then increase the field in 
small steps till all spins point along the positive $x$-axis. 
At each step, the field is held fixed while the system 
relaxes to a fixed point configuration under the dynamics 
considered above. This yields a line of fixed points. The 
graph of magnetization of fixed point configurations versus 
the applied field gives the magnetization curve in increasing 
field. Magnetization in decreasing field is obtained 
similarly. If the magnetization in decreasing field follows a 
different path than the one in increasing field, the system 
is said to show hysteresis i.e. history-dependent effects.

We wish to know if the system characterized by Hamiltonian 
(1) exhibits hysteresis, and if so what is the shape of the 
hysteresis loop. Another question of interest is whether 
there is a critical value of disorder that qualitatively 
separates the hysteretic response of weakly disordered 
systems from that of strongly disordered systems. The meaning 
of critical disorder in this context is best explained by a 
reference to earlier work of Sethna et al~\cite{sethna1} on 
disorder-driven hysteresis in the random field Ising model. 
They consider a Hamiltonian similar to (1) but in their case 
the spins and the fields $h$ and $h_i$ are scalar quantities; 
spins take the values $\pm1$, and $h_i$ is a random variable 
chosen from a Gaussian distribution centered at zero and 
having variance equal to $\sigma^2$. Their results are based 
on a combination of numerical simulations and analysis, but 
are quite intuitive as well. These may be summarized as 
follows. In the limit $\sigma \rightarrow 0$, as the applied 
field is increased from $h=-\infty$ to $h=\infty$, each spin 
and therefore the magnetization per site flips up from $-1$ 
to $+1$ at $h=zJ$ where $z$ is the number of nearest 
neighbors on the lattice.  As $\sigma$ is increased, the size 
of the jump in the magnetization decreases and eventually 
vanishes at $h=h_c$ if $\sigma = \sigma_c$. For $\sigma > 
\sigma_c$, the magnetization becomes a smooth function of the 
applied field.  The point $\{h=h_c$, $\sigma = \sigma_c\}$ is 
a non-equilibrium critical point characterized by diverging 
correlation length and scaling laws reminiscent of 
equilibrium critical phenomena. The parameter $\sigma$ 
measures the width of the random field distribution and 
therefore the amount of disorder in the system. The disorder 
is said to be critical if $\sigma=\sigma_c$. The 
non-equilibrium critical point may also be studied by fixing 
the disorder in the system, say by setting $\sigma=1$ and 
tuning the exchange interaction $J$ and the applied field $h$ 
to the critical point $\{h=h_c,J=J_c\}$. Now the 
magnetization curves in increasing and decreasing fields 
would be smooth for $J<J_c$, but discontinuous for $J>J_c$. 
The size of the discontinuity would go to zero as $J$ 
approaches $J_c$ from above.

The question is if there is a critical value $J_c$ as we go 
from scalar to vector spins?  In the random field Ising model 
the spins have the value $+1$ or $-1$. Therefore the 
boundaries between domains of positive and negative 
magnetization are sharp. The width of the domain wall is 
equal to the distance between nearest neighbors on the 
lattice.  Vector spins can continuously change their 
orientation from one domain to another over arbitrarily thick 
domain walls. Phase transition in a system depends on the 
balance between energy gained by forming a large domain, and 
energy lost in having to protect it by a domain wall. The 
energetics of this competition in continuous spins is very 
different from that in Ising spins ~\cite{imry-ma}. It shows 
that continuous spins in random fields cannot acquire a 
spontaneous long range order below four dimensions, while the 
lower critical dimension for Ising spins is two. This means 
that the critical hysteresis observed in the random field 
Ising model in three dimensions may disappear when we go over 
to vector spins. Although the focus of our work is on the 
shapes of hysteresis rather than criticality, we shall return 
to this point after presenting our results.

It is useful to have a brief preview of our results before 
getting into details. It also gives us an opportunity to 
mention some unusual aspects of hysteresis in continuous spin 
systems. In our model, the disorder has a fixed magnitude and 
sets the energy scale of the system. The behavior of the 
model is therefore determined by the parameter $J$. If $J=0$, 
the spins decouple and there can be no hysteresis in the zero 
frequency limit of the driving field. We find that the 
behavior of the model for small values of $J$ is 
qualitatively similar to the behavior for $J=0$. This is true 
in the mean field analysis as well as numerical simulations 
of the model on a lattice with nearest neighbor interactions. 
For large values of $J$ we may expect hysteresis as well as 
jumps in the magnetization. The basis for this expectation is 
the following. Large $J$ means relatively weak disorder. Thus 
the spins are mostly aligned parallel to each other. As the 
applied field is swept from $h=-\infty$ to $h=\infty$, we 
expect the majority of spins to reverse their direction at a 
critical field $h=h_c$. The field $h_c$ is determined by the 
energy required to flip the least stable spin in the system 
that triggers a large avalanche of flipped spins. For 
discrete Ising spins with $z$ nearest neighbors, $h_c$ is of 
the order of $zJ$ in the limit of weak disorder. However, in 
the case of continuous spins, the least stable spin can 
reverse itself by rotating smoothly along with its neighbors. 
In other words, the energy barrier for magnetization reversal 
may be zero for continuous spins in the strong coupling limit 
just as it is in the weak coupling limit. We find that the 
mean field theory predicts a non-zero value for $h_c$ but 
simulations based on short range interactions on a lattice 
indicate $h_c \rightarrow 0$ in the limit $J \rightarrow 
\infty$.

Hysteresis in continuous spin systems at intermediate values 
of $J$ where order and disorder compete with each other has 
several unusual features. Normally if a system shows 
hysteresis, the magnetization curves in increasing and 
decreasing fields are separated by the widest margin in the 
middle at $h=0$. We find that there is a range of $J$ values 
where the magnetization curves for the XY model in the mean 
field approximation overlap each other in the middle but 
split from each other as we go away from $h=0$ in either 
direction. Numerical simulations of the XY model show a 
qualitatively similar behavior although there are significant 
differences between simulations and the predictions of the 
mean field theory. Broadly speaking, discontinuities in the 
magnetization curves predicted by the mean field theory 
appear to be absent in simulations. The mean field theory of 
hysteresis in the Heisenberg model has an unusual feature as 
well. Usually the mean field solution is determined by the 
intersection of a straight line with an $S$-shaped curve. In 
this case the mid portion of the $S$-shaped curve is a 
straight line itself. This gives rise to some interesting 
effects that are seen in corresponding simulations as well. 
In the following, we examine these issues in detail.

\subsection{XY model}

It is instructive to look at the mean field dynamics of the 
XY model numerically before presenting the analytic solution. 
Let us set the applied field equal to zero ($h=0$), start 
with an arbitrary initial state characterized by 
magnetization $m_0$, and iterate equation (9) until a fixed 
point is reached. The results are shown in figure (2). We 
find two critical values of $J$: $J_{c1}\approx 1.489$, and 
$J_{c2} =2$. These values characterize discontinuities in the 
fixed point behavior in increasing and decreasing $J$ 
respectively as described below.

The blue curve in figure (2) shows magnetization of fixed 
points of equation (9) for increasing $J$. We start with 
$J=0$, and increase $J$ in small steps of $\Delta J$. At each 
value of $J$, the magnetization $m(J-\Delta J)$ of the 
previous fixed point is used as an starting point for the 
iteration of equations. The precise value of $\Delta J$ is 
unimportant. We have chosen a value of $\Delta J$ that is 
small enough so that the line of fixed points appears as a 
continuous curve on the scale of figure (2). For increasing 
$J$, the fixed point magnetization $m$ is zero in the range 
$0 \le J < J_{c2}$. At $J=J_{c2}$, it jumps to $m \approx 
.92$, and follows the blue curve as $J$ is increased further. 
The return path in decreasing $J$ is identical with the blue 
curve up to $J \le \ J_{c2}$, but there is no discontinuity 
in the return path at $J=J_{c2}$. It continues smoothly along 
the green curve up to $J = J_{c1}$ at which point it jumps 
down to zero and remains zero for $0 \le J < J_{c1} $. The 
red curve shows a set of unstable fixed points in the range 
$J_{c1} \le J \le J_{c2}$. An unstable fixed point is not 
realized under iterations of equation (9) because its domain 
of attraction is zero. However, for a fixed $J$, the 
magnetization $m_u$ of the unstable fixed point separates the 
domains of attraction of the two stable fixed points at the 
same value of $J$. If the magnetization of the starting state 
$m_0$ is less than $m_u$, the equations iterate to the fixed 
point associated with increasing $J$. If $m_0 > m_u$ the 
equations iterate to the corresponding fixed point for 
decreasing $J$. The reason for the existence of two stable 
and one unstable fixed point in the range $J_{c1} \le J \le 
J_{c2}$ may be understood analytically as follows. Let us 
define,

\be f(u_t) = \left. \frac{1}{2\pi}\int_0^{2\pi} \frac{u_t + 
\cos \alpha_i}{[ 1+2 u_t \cos \alpha_i + u_t^2 
]^{\frac{1}{2}}} d \alpha_i \right. \mbox{ where } u_t=J_0 
m(t) +h \ee

The quantity $f(u_t)$ can be written in terms of complete 
elliptic integrals of the first and second 
kinds~\cite{mirollo}:

\be f(u_t)=\frac{1}{\pi u_t} \left[ (u_t-1) K \left( 
\frac{2\sqrt{u_t}}{1+u_t}\right) + (u_t+1) E \left( 
\frac{2\sqrt{u_t}}{1+u_t}\right) \right] \ee

The red curve in figure (3) shows a graph of $f(u)$ vs $u$. 
It is a continuous, increasing function of $u$ in the range 
$0 \le u \le \infty$. Some special values are: $f(0)=0; 
f(1)=\frac{2}{\pi}; f(\infty)=1; f^{\prime}(0)=\frac{1}{2}; 
\lim_{u \rightarrow 1} f^{\prime}(u)=\infty$. Using equation 
(12), equation (9) may be rewritten as,

\be u_{t+1} = J_0 f(u_t) + h \ee Fixed points of equation 
(14) are the roots of the equation, \be \frac{u-h}{J_0} = 
f(u) ;\mbox{ where } \lim_{t \rightarrow \infty} u_t=u =J_0 
m+h; \mbox{ and } \lim_{t\rightarrow\infty}m(t)=m \ee

The roots of equation (15) are determined by the intersection 
of the curve $f(u)$ with the straight line $(u-h)/J_0$. For 
$h=0$, $u=0$ is always a root because $f(0)=0$, and the 
straight line $u/J_0$ passes through origin. However, 
equation (15) may have up to three more roots because $f(u)$ 
is an S-shaped curve, and a straight line with appropriate 
slope may cut it at three points. Consider straight lines 
passing through the origin and having decreasing slopes i.e. 
lines $u/J_0$ with increasing $J_0$. Let $J_{c1}$ and 
$J_{c2}$ be the smallest and the largest values of $J_0$ 
respectively at which the line $u/J_0$ meets the S-shaped 
curve $f(u)$ tangentially as shown in figure (3). As 
mentioned at the beginning of this section, $J_{c1} \approx 
1.489$ and $J_{c2} =2$. There are three non-zero roots of 
equation (15) in the range $J_{c1} \le J_0 \le J_{c2}$, and 
only one non-zero root for $J_0 > J_{c2}$. Which of these 
roots is actually realized by the dynamics is determined by 
the starting point $u_{0}$ used in iterating equation (14). 
The stability of a root can be checked by analyzing equation 
(14) in the neighborhood of its fixed point ~\cite{mirollo}. 
However, the full equation is necessary to determine the 
domain of attraction of a stable fixed point.

Next we consider equation (14) for a fixed value of $J_0$ but 
in a varying field $h$. Starting from a sufficiently negative 
field $h=h_{min}$ where the stable configuration has most 
spins pointing along the negative $x$-axis, the field is 
increased in small steps $\Delta h$ to $h=h_{max}$ where most 
spins point along the positive $x$-axis. Figure 4 shows the 
fixed point magnetization $m$ as the field is increased from 
$h_{min}=-1.5$ to $h_{max}=1.5$ and back to $h_{min}=-1.5$ in 
steps of size $\Delta h=.01$. Data for three representative 
values of $J$ are shown: $J_0= 2$ (red), $J_0=1.25$ (green), 
and $J_0=.25$ (blue). $J=2$ shows a familiar looking 
hysteresis loop but $J=1.25$ shows a somewhat unfamiliar 
behavior. In this case, there are two symmetrically placed 
windows of positive and negative applied fields where the 
system shows hysteresis but there is no hysteresis in the 
intermediate region near zero applied field. For $J=.25$ 
there is no discernible hysteresis on the scale of figure 
(4).

The variety of behavior seen in figure (4) may be understood 
as follows.  We saw in figure (3) that spontaneous 
magnetization is possible only if $J_0 > J_{c1}\approx 
1.498$. Spontaneous magnetization in zero applied field gives 
rise to the possibility of hysteresis as the applied field is 
cycled up and down across the value $h=0$. Therefore the 
hysteresis loop for $J_0=2$ (red curve) in figure (4) 
centered around $h=0$ is to be expected. We do not expect the 
green curve ($J_0=1.25$), or the blue curve($J_0=.25$) to 
show a hysteresis at $h=0$. This is born out by figure (4). 
What is surprising at first sight is that the green curve in 
figure (4) shows two small hysteresis loops in applied fields 
centered around $h \approx \pm0.2$. We can understand this 
with the help of figure (5) that shows three straight lines 
$(u-h)/J_0$ for $J_0=1.25$; and $h=.1, .2$, and $.3$ 
respectively. These lines are superimposed on the graph of 
$f(u)$ for $u \ge 0$. The two lines corresponding to $h=.1$ 
and $h=.3$ cut $f(u)$ only once. The point of intersection 
corresponds to a stable fixed point. There is only one stable 
fixed point at applied fields $h=.1$, and $h=.3$. Thus the 
magnetization at $h=.1$ and $h=.3$ has the same value whether 
the applied field is increasing or decreasing. This explains 
why the green curve shows no hysteresis in the vicinity of $h 
\approx 0.1$ and $h \approx 0.3$. However, the straight line 
corresponding to $h=.2$ cuts $f(u)$ at three points. Two of 
these points are stable fixed points: one in increasing 
applied field and the other in decreasing field. The third 
non-zero fixed point is an unstable fixed point. This gives 
rise to hysteresis in a small window of applied field 
centered around $h \approx 0.2$. By symmetry there is a 
similar window of hysteresis around $h \approx -0.2$.

\subsection{Heisenberg model}

Making a transformation of variables $\mu_i=\cos \alpha_i$, 
and $u_t=J_0 m(t) + h$, equation (11) may be rewritten as:

\bea u_{t+1}=J_0 g(u_t)+h, \mbox{ where }\nonumber \\ g(u_t) 
=\left. \frac{1}{2}\int_{-1}^{1} \frac{u_t + \mu_i}{[ 1+2 u_t 
\mu_i + u_t^2
]^{\frac{1}{2}}} d\mu_i \right. \eea

The integral in equation (16) is easily evaluated and yields

\bea g(u_t)= \frac{2}{3}u_t \mbox{ if } |u_t| \le 1 \nonumber 
\\ g(u_t)= 1 - \frac{1}{3 u_t^2} \mbox{ if } u_t > 1 
\nonumber \\ g(u_t)= -1 + \frac{1}{3 u_t^2} \mbox{ if } u_t < 
-1 \eea

Figure (6) shows a graph of $g(u)$ with the line $2u/3$ 
superimposed on it. The fixed points of the iterative 
equation are determined by the equation $m=g(J_0 m +h)$. For 
$h=0$, and $J_0 |m| \le 1$, the fixed point is determined by 
the equation $m=\frac{2}{3}J_0 m$. If $J_0=\frac{3}{2}$, then 
any value of $m$ in the range $-\frac{2}{3} \le m \le 
\frac{2}{3}$ satisfies the fixed point equation. This is 
rather unusual in a mean field theory. Normally, the 
spontaneous magnetization in a mean field theory is 
determined by the intersection of a straight line with an 
S-shaped curve. In the present case the S-shaped curve is 
itself a straight line in the interval $-1 \le J_0 m \le 1$. 
This means that in the absence of an applied field, the zero 
temperature magnetization in the random field Heisenberg 
model is zero if $J_0 < \frac{3}{2}$, and can have an 
arbitrary value in the range $-\frac{2}{3} \le m \le 
\frac{2}{3}$ if $J_0=\frac{3}{2}$. For $J_0 > \frac{3}{2}$, 
$|m| > \frac{2}{3}$ and increases with applied field $h$. 
Figure (7) shows the magnetization curves in a cyclic field 
(varying infinitely slowly in the sense explained earlier) 
for $J_0=1$(pink), $J_0=1.5$(blue), $J_0=1.75$(green), and 
$J_0=2$(red). As expected from the above analysis, there is 
no hysteresis in the case $J_0=1$ and $J_0=1.5$, although in 
the case $J_0=1.5$ the magnetization shows a finite jump at 
$h=0$. There is hysteresis for $J_0=1.75$ and $J_0=2$ with 
the area of the hysteresis loop increasing with $J_0$.

\subsection{Peculiar criticality}

In this section we attempt to place our calculations in the 
context of extant work on critical hysteresis in 
$n$-component vector spin systems with quenched disorder. The 
extant work employs soft continuous spins and a Gaussian 
distribution of quenched field, while we have used hard 
continuous spins and random fields in the form of randomly 
oriented unit vectors. Vector spins with $n\ge2$ are called 
continuous spins. These can be hard or soft. Hard continuous 
spins have a fixed length but can make any angle from a 
reference axis. Computer simulations commonly use hard spins 
on a lattice. We have used hard spins for numerical as well 
as analytic work. Soft spins are continuous in angle as well 
as magnitude. Momentum space renormalization group uses soft 
spins ~\cite{sethna1,silveira1}. It usually starts out with 
hard spins on a lattice but transforms them into soft spins 
that can take any real value but are constrained to remain 
close to a fixed length. This is done by introducing an 
effective on-site potential. Similarly continuum limit of the 
lattice is taken but a cutoff on the maximum wave-vector is 
introduced. There is some evidence that the critical behavior 
of models in the renormalization group theory is independent 
of the additional parameters introduced by the effective 
on-site potential and momentum cutoff
~\cite{shukla}. However, it is not independent of the form of
the random field distribution.

Hartmann et al ~\cite{hartmann} have shown numerically that 
the critical exponents of three dimensional random field 
Ising model with Gaussian distribution of random fields are 
significantly different from those of the same model with a 
bimodal distribution of random fields. Analytic results in 
three dimensions are not available. What is available is a 
perturbation series for critical exponents in $6-\epsilon$ 
dimensions for a Gaussian distribution of the quenched field
~\cite{silveira1}. It shows that similar to the random field 
~Ising ($n$=1) model \cite{sethna1}, random field XY ($n$=2) 
~and Heisenberg ($n$=3) models have a critical point on the
hysteresis loop. The critical exponents depend on $n$ below 6 
dimensions, but are independent of $n$ in 6 and higher 
dimensions. The significance of 6 dimensions is that it is 
equal to the upper critical dimension of the model with a 
Gaussian distribution of the random field ~\cite{parisi}. In 
6 and higher dimensions the action is adequately described by 
a quadratic term, and higher order terms become irrelevant in 
the renormalization group sense. The quadratic action can be 
solved exactly, and the solution is often called the mean 
field solution (presumably because it gives the same critical 
exponents as a mean field solution based on infinite range 
interactions). In this variant of the mean field theory, the 
hysteresis loops for $n$=1, 2, and 3 exhibit a critical point 
as the width of the Gaussian disorder increases, but the 
critical exponents do not depend upon $n$.

In our variant of the mean field theory based on infinite 
range interactions, the critical behavior for $n=2$ is 
different from that of $n=3$. This should not raise a serious 
concern because we use a different distribution of the random 
field than used in reference ~\cite{silveira1}. Our random 
field distribution for $n$= 2 and 3 is a continuous analog of 
bimodal distribution in the case $n$=1. For $n=1$, we know 
that Gaussian and bimodal distributions of random field give 
two different sets of critical exponents in 3 dimensions. 
This difference may persist even above the upper critical 
dimension although to our knowledge the upper critical 
dimension for a bimodal distribution is not known precisely. 
Nevertheless the striking difference between the nature of 
criticality for $n=2$ and $n=3$ in our mean field theory is 
interesting and could not have been anticipated beforehand. 
We may therefore take a closer look at the algebraic 
mechanism producing this difference and also the difference 
from the mean field theory of the random field Ising model 
based on infinite range interactions and a Gaussian 
distribution of the random field. In these exactly solved 
cases the equations of motion have a similar form but the 
signs of various terms depend on the details of the model. 
This produces distinct critical behavior in each case. We 
have,

\be u_{t+1} = J_0 \mbox{ Erf}\left( 
\frac{u_t}{\sqrt{2}}\right) + h \mbox{\hspace{1cm} ( Ising 
model; Gaussian random field )}\nonumber \ee \be u_{t+1} = 
J_0 f(u_t) + h\mbox{\hspace{1cm} ( XY model; fixed 
magnitude/random orientation field )} \nonumber \ee \be 
u_{t+1}=J_0 g(u_t)+h \mbox{\hspace{1cm} ( Heisenberg model; 
fixed magnitude/random orientation field )} \nonumber \ee

where $u_t=J_0 m(t) +h$. The first equation is for the random 
field Ising model with a Gaussian distribution centered at 
zero and having unit variance. It can be easily derived and 
at its fixed point it reduces to the mean field equation 
studied by Sethna et al ~\cite{sethna1}. The equations for 
the XY and Heisenberg models were derived in sections III A 
and III B where the functions $f(u)$ and $g(u)$ are also 
defined. For simplicity, let us set $h=0$ and confine to 
$u\ge0$. Now suppose we start with a small value of $u_t$ and 
iterate the above equations of motion till we reach a fixed 
point $u^*$. We focus on the behavior of $u^*$ as a function 
of $J$. In each case, there is a threshold $J_c$ such that 
$u^*=0$ for $J < J_c$. We get $J_c$ = $\sqrt{\frac{\pi}{2}}$, 
2, and $\frac{3}{2}$ for $n$=1, 2, and 3 respectively. At 
$J=J_c$, there is a transition to a non-zero value of $u^*$. 
This transition is continuous for $n=1$, discontinuous for 
$n=2$, and peculiarly discontinuous for $n=3$ in the sense 
that $u^*$ can have any value in the range $0 \le u^* \le 1$. 
Thus the transitions for $n$ = 1, 2, and 3 are distinct from 
each other.

It is not difficult to understand the above results 
analytically. The functions Erf$(u)$, $f(u)$, and $g(u)$ are 
all zero at $u=0$ and their first derivatives with respect to 
$u$ are positive. Erf$(u)$ is concave down for $u\ge0$; 
$f(u)$ is concave up for $0\le u<1$ and concave down for 
$u>1$; $g(u)$ has zero curvature for $0\le u<1$ and concave 
down for $u>1$. It is also instructive to write the leading 
terms in the series expansion of the right hand side. We get 
the following expressions for the Ising, XY, and Heisenberg 
spins respectively,

\be n=1: \hspace{1cm} u_{t+1} = \sqrt{\frac{2}{\pi}} J_0 
u_t-\frac{J_0}{3\sqrt{2\pi}} u_t^3 + \ldots ( u_t 
\longrightarrow 0);\hspace{1cm} u_{t+1} = J_0 - 
\frac{J_0}{\sqrt{\pi} u_t}\exp{\left(-\frac{u_t^2}{2}\right)}
+ \ldots ( u_t \longrightarrow \infty) \nonumber \ee

\be n=2: \hspace{1.5cm} u_{t+1} = 
\frac{J_0}{2}u_t+\frac{J_0}{16} u_t^3
+ \ldots ( 0\le u_t\le 1);\hspace{1cm} u_{t+1}= J_0 - 
  \frac{J_0}{4 u_t^2} - \ldots ( u_t>1)\nonumber \ee

\be n=3: \hspace{1.5cm} u_{t+1} = \frac{2 J_0}{3}u_t ( 0\le 
u_t\le 1);\hspace{1cm} u_{t+1}= J_0 - \frac{J_0}{3 u_t^2} 
\hspace{1cm}( u_t
>1)\nonumber \ee

The leading terms of the recursion relations in the limit 
$u_t \rightarrow 0$ show that the fixed point $u^* 
\rightarrow 0$ if $\sqrt{\frac{2}{\pi}}J_0 < 1$ for $n=1$, 
$\frac{J_0}{2}<1$ for $n=2$, and $\frac{2 J_0}{3}<1$ for 
$n=3$. This yields the critical values $J_c$ mentioned above. 
For $n=1$ and $J>J_c$, $u^* \approx (J-J_c)^{1/2}$ i.e. it 
has a square root singularity characteristic of mean field 
critical behavior. In the case $n=2$, the cubic term has a 
positive sign and therefore a physically acceptable solution 
does not grow continuously from $u^*=0$ at $J=J_c$. The 
solution of the full equation shows that $u^*$ has a first 
order jump in this case. The recursion relation for $n=3$ is 
peculiar because it does not have any non-linear terms. At 
$J_c=3/2$ any value of $u^*$ ($0 \le u^* \le 1$) satisfies 
the fixed point equation. For $J>J_c$, $u^*$ increases with 
$J$ but remains bounded below $J$.

\subsection{Simulations}

Figure (8) and figure (9) show magnetization curves for the 
random field XY and Heisenberg models respectively in a 
slowly varying cyclic field. The data is obtained from 
simulation of the model on a simple cubic (sc) lattice with 
nearest neighbor (nn) interactions. In order to keep the 
computer time within reasonable limits, the XY model is 
simulated on a lattice of size $100^3$, and the Heisenberg 
model on a lattice of size $50^3$ with periodic boundary 
conditions. Graphs are presented for various values of $J$ as 
indicated in the captions for the figures. For each value of 
$J$, the applied field is cycled in small steps between two 
large values that saturate the magnetization along negative 
and positive $x$-axis respectively. For clarity, the figures 
depict only a part of the simulation data in a small range of 
the applied field where variation in magnetization is most 
pronounced. At each step of the applied field the system is 
allowed to relax till it reaches a fixed point. We assume 
that the system has reached a fixed point if the projection 
of each spin along $x$-axis remains invariant within an error 
of $10^{-5}$.

The mean field theory predicts the absence of hysteresis in the $XY$ 
model if $J_0<1.498$. The energy scale in our model is set by the 
disorder term. Thus the behavior of the mean field model at $J_0$ may be 
compared with the behavior of the nn model on a sc lattice at $6J$. As 
an order of magnitude estimate, we expect the absence of hysteresis on a 
sc lattice if $6J<1.498$ or $J<.25$ approximately. This is qualitatively 
in accordance with the result of simulations shown in figure (8). The 
magnetization curves for $J=.1$ show no discernible hysteresis on the 
scale of the figure. At $J=.2$, we find two isolated hysteresis loops 
separated by a region of zero hysteresis near $h=0$. This is 
qualitatively similar to the prediction of the mean field theory. With 
increasing $J$ the two isolated loops widen, gradually merge with each 
other, and the overall shape of the hysteresis loop evolves as indicated 
in figure (8). For much larger values of $J$ the hysteresis loop becomes 
narrower and more vertical. Within numerical errors, magnetization 
curves in increasing and decreasing fields approach a step function at 
$h=0$, and hysteresis appears to vanish for $J \ge 1$. The large $J$ 
regime marks a qualitative difference between the prediction of the mean 
field theory and the simulations. The mean field theory predicts 
hysteresis but simulations on cubic lattices with nearest neighbor 
interactions show no hysteresis. This discrepancy may be attributed to 
the use of infinite range interactions in the mean field theory. The 
energy barrier for rotation of a strategically placed spin may be 
significantly smaller if its nearest neighbors alone are taken into 
account rather than all spins in the system. The dynamics based on nn 
interactions initiates a rotation at the least stable site and gradually 
spreads it on adjacent sites in the neighborhood. Large $J$ simulations 
take an enormously long time to reach a fixed point in the neighborhood 
of $h=0$, but the end result appears to be simply a reversal of 
saturation magnetization when the sign of $h$ is reversed. In the limit 
$J \rightarrow \infty$, the system effectively acts as a single spin 
having the total magnetization of the system. Just as an isolated spin 
in the limit $J=0$ does not show any hysteresis so also the entire 
system in the limit $J \rightarrow \infty$. The main difference between 
the magnetization curves in the limits $J \rightarrow 0$ and $J 
\rightarrow \infty$ lies in their shape, but this is understandable if 
we rescale the applied field appropriately with the total magnetization 
of the system.

Simulations of the Heisenberg model are also in reasonable 
agreement with the predictions of the mean field theory 
except for large values of $J$. The mean field theory 
predicts hysteresis if $J_0> 3/2$. This corresponds to 
$J>.25$ approximately.  Simulations do not show any 
significant hysteresis if $J \le.25$. Figure (9) shows a 
magnetization curve for $J=.25$ that reverses itself when the 
field is reversed. The magnetization is linear in the applied 
field over a wide range around $h=0$. This is in qualitative 
agreement with the prediction of the mean field theory. 
Simulations for $J=.4$ and $J=.5$ show typical hysteresis 
loops although the range of applied field over which 
perceptible hysteresis is observed is an order of magnitude 
smaller than the range predicted by the mean field theory. 
The main difference between the simulations and the mean 
field theory lies at large values of $J$. The magnetization 
curves shown in figure (9) for $J=1$ appear to form a narrow 
nearly vertical hysteresis loop. However size of the steps 
used to increase and decrease the applied field in figure (9) 
is of the order of the width of the apparent hysteresis loop. 
Simulations based on smaller steps and higher accuracy in 
determining the fixed points suggest that the hysteresis loop 
vanishes for $J \ge 1$ and the magnetization has a 
first-order jump at $h=0$.

\section{Concluding remarks}

We have analyzed a simple model to study the effect of 
quenched disorder on hysteresis in magnetic systems of 
continuous symmetry. The model is obtained by adding quenched 
disorder and zero temperature dynamics to the well 
established XY and Heisenberg models of ferromagnetism. The 
quenched disorder is in the form of randomly oriented fields 
of unit magnitude. Is this model applicable to experiments? 
We have argued that thermal fluctuations are of secondary 
importance in disorder-driven hysteresis. Therefore the use 
of zero temperature dynamics may not be serious. It has the 
virtue of being deterministic and therefore easier to analyze 
theoretically. A large number of studies on disordered 
systems employ zero temperature dynamics for these reasons. 
Randomly oriented crystal fields are also not uncommon in 
amorphous materials. These are dipolar or quadrupolar but if 
the activation barriers are large, may act like quenched 
random fields as a spin or domain pointing one way gets hard 
to dislodge. Thus the basic ingredients of our model are 
chosen to make a minimal model for understanding experiments.  
The parameters of the resulting model are: $n$ components of 
vector spins, exchange interaction $J$, and the applied field 
$h$. Effectively, there are just two parameters; integer $n$ 
and real $J$. This is because the middle term in Hamiltonian 
(1) does not have a tunable value, and the field $h$ is 
cycled between $-\infty$ and $\infty$. A two parameter model 
may not capture details of hysteresis in various materials 
but it provides a caricature of experimental observations. 
The variety of shapes of hysteresis loops are particularly 
striking for the XY model ($n$=2). As $J$ is varied, we get 
familiar as well as rather unusual shapes of loops. The 
unusual shapes have been noted earlier in magnetic and other 
materials. These are known as wasp-waisted ~\cite{bennett} or 
double-flag shaped loops ~\cite{cardone}. These shapes have a 
kind of weak universality in the sense that they are seen in 
the mean field theory, simulations on three dimensional 
lattices, and experiments in diverse systems. Similar shapes 
are also seen in the random field Blume-Emery-Griffiths model 
and other models of plastic depinning of driven disordered 
systems.

Soft continuous spins with Gaussian random fields have been 
used earlier to study critical hysteresis in $6-\epsilon$ 
dimensions in the renormalization group theory.  Where does 
our mean field calculation sit in this context? We note that 
our mean field results do not match the renormalization group 
results in any limit. There are two possible reasons for 
this. Firstly, we have used a different distribution of 
random fields than the one used in $6-\epsilon$ expansion. 
The form of random field distribution appears to be important 
in determining critical hysteresis. Secondly, we do not have 
the benefit of an appropriate renormalization group study of 
our variant of the model, nor do we know the upper critical 
dimension of our model precisely. Before the renormalization 
group theory, mean field theory was viewed as an approximate 
but self consistent theory of critical behavior in 3 
dimensions because it neglected fluctuations. This variant of 
mean field theory was based on infinitely weak but long 
ranged interactions in the system. It represented the effect 
of the entire system on an individual spin by an effective 
field while keeping the energy of the system extensive. The 
renormalization group has given another connotation to mean 
field theory. In its framework, the mean field theory becomes 
a reduced theory based on a quadratic action that is exact at 
and above an upper critical dimension where fluctuations are 
negligible. The upper critical dimension for pure 
(non-disordered) magnetic systems is 4, and in this case the 
two variants of the mean field theory predict the same 
critical behavior. This is understandable because the 
effective field is proportional to the order parameter and 
the effective action is therefore quadratic. The upper 
critical dimension for a disordered spin model with a 
Gaussian random field is 6. In this case also, an explicit 
calculation for the random field Ising model with Gaussian 
field shows that the two variants of mean field theory yield 
the same critical behavior. However, when the randomness is 
of the form of randomly oriented unit vectors, we have 
analyzed only one variant of mean field theory that is based 
on infinitely weak but long range forces. It predicts 
strikingly different critical behavior in XY and Heisenberg 
models respectively. The case $n=2$ has a first order 
transition, and $n=3$ an unusual transition as discussed in 
section III C. A somewhat similar case of first as well as 
second order depinning transition in the mean field theory 
occurs in a viscoelasic model of driven disordered systems
~\cite{marchetti,saunders}. Thus we have a number of model
specific results. Evidently more work is required to make any 
general connection between the random field distribution and 
the nature of criticality in the model, and to connect a 
conventional mean field theory to a limiting form of 
renormalization group theory above an upper critical 
dimension.

The new framework for understanding critical behavior also 
uses the idea of a lower critical dimension. Below the lower 
critical dimension the fluctuations are so great that the 
system does not order at all and therefore there is no 
question of a phase transition. The lower critical dimension 
for an equilibrium transition in an $n$-component spin system 
in a Gaussian random field is 2 for $n=1$, and 4 for $n \ge 
2$. To our knowledge, the lower critical dimension for the 
case of randomly oriented unit vectors is not known. However, 
we mention a few issues that may bear on experiments in three 
dimensions irrespective of the form of randomness 
characterizing the system. It has been argued that critical 
hysteresis in a Gaussian random field Ising model is in the 
same universality class as the corresponding equilibrium 
critical point ~\cite{sethna1,silveira1}. There is a 
reasonable experimental evidence for this ~\cite{sethna1}. In 
the case of Gaussian random field XY and Heisenberg models, 
the lower critical dimension lies above 3. Does it 
necessarily mean the absence of critical hysteresis in these 
models in 3 dimensions? The situation is not entirely 
convincing either theoretically or experimentally. The 
argument for a lower critical dimension is based on 
spontaneous symmetry breaking in the absence of an applied 
field. If the critical point on the hysteresis loop were to 
occur at a non-zero value of magnetization or the applied 
field then a unique direction is already chosen by the 
corresponding magnetization or the applied field. In this 
case we may observe critical hysteresis in 3 dimensions with 
critical exponents appropriate for the Gaussian random field 
Ising model. Much of critical hysteresis seen in experiments 
may belong to this case but there is discrepancy between some 
experiments and theory ~\cite{silveira2}. It may be that 
quenched disorder in experimental systems is not 
characterized adequately by Gaussian random fields or 
randomly oriented unit vectors. The presence of demagnetizing 
fields and dipolar forces in materials used for experiments 
are likely to change the simple theoretical picture based on 
on-site random field disorder. This applies equally to mean 
field theory and the renormalization group approach based on 
expansions around a quadratic action.

In the absence of exact solutions in three dimensions, 
simulations of models may be more relevant to experiments. 
Our simulations produce rather smooth hysteresis loops for 
small and moderate values of $J$ suggesting the absence of 
jumps in the magnetization. Phase transitions in complex 
systems are difficult to decide on the basis of numerical 
work alone, and therefore we have focused on the shape of 
hysteresis loops. The shapes are not universal but this does 
not diminish their importance in the application of magnetic 
materials. The relationship between the shape of hysteresis 
loops and the defect mediated process of magnetization 
reversal is also interesting. This has been studied at zero 
temperature numerically in a two-dimensional XY-model with 
weak random anisotropy ~\cite{dieny}. We hope future studies 
on these lines will clarify the relationship of the 
hysteresis loops to the underlying energy landscape as well 
as the patterns of spin configurations such as vortex loops 
in three dimensional XY model.

\begin{acknowledgments} PS thanks the School of Mathematics, 
University of Southampton for hospitality during a short 
visit funded by the Royal Society when the work presented 
here was started. He also thanks T J Sluckin for discussions 
in the initial stage of this work, and Deepak Dhar for a 
critical reading of the manuscript. \end{acknowledgments}

\begin{figure}[p]

\includegraphics[width=.75\textwidth,angle=0]{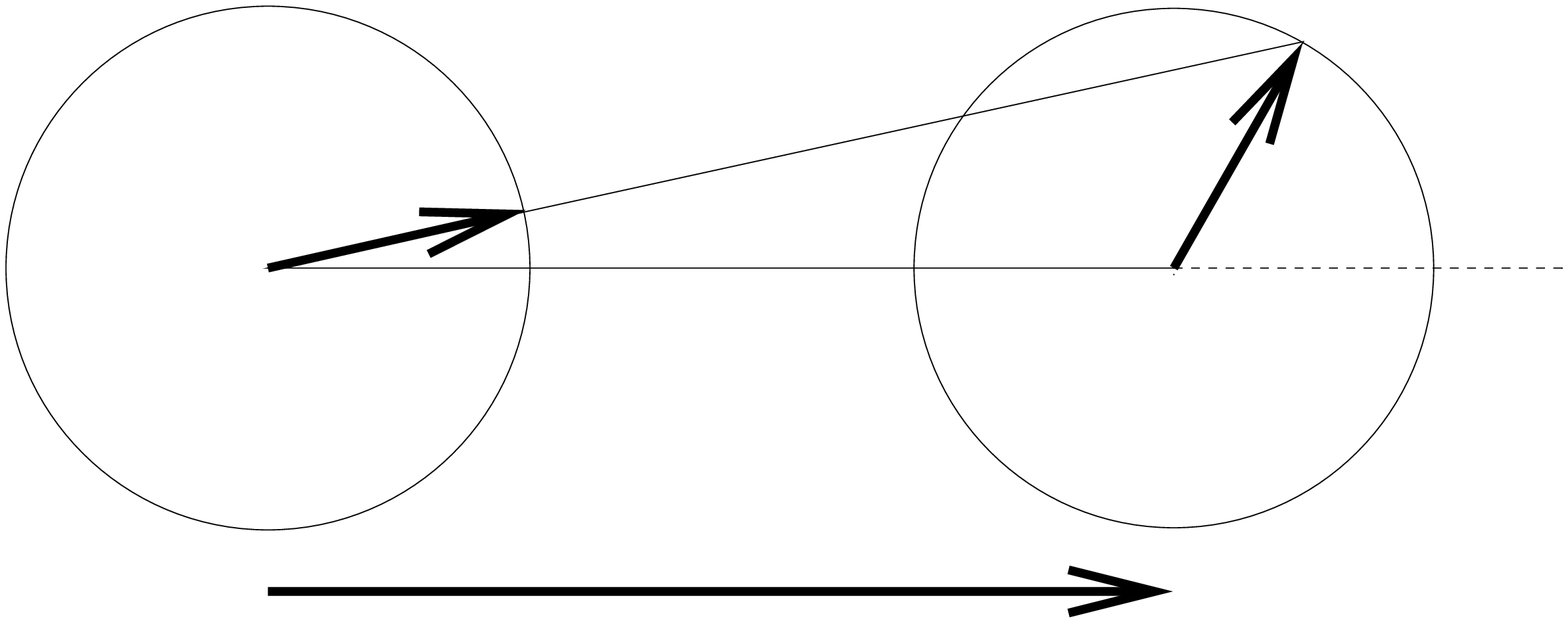}

{{\vspace{-3.75cm}{{\Large\bf{\hspace{3.25cm}$\theta_i(t+1)$ 
\hspace{3cm} $\alpha_i$}}}}}

\vspace{2cm}

{{\huge\bf{{$J_0 m(t) +h$}}}}

\vspace{1cm}

\caption{ A geometrical representation of the dynamics of 
random field XY model showing a vector relationship between 
the updated spin $\vec{S}_i(t+1)$, the random field 
$\vec{h}_i$, and the mean field $J_0 m(t) +h$ at site $i$.} 
\label{fig1} \end{figure}

\begin{figure}[p] 
\includegraphics[width=.75\textwidth,angle=0]{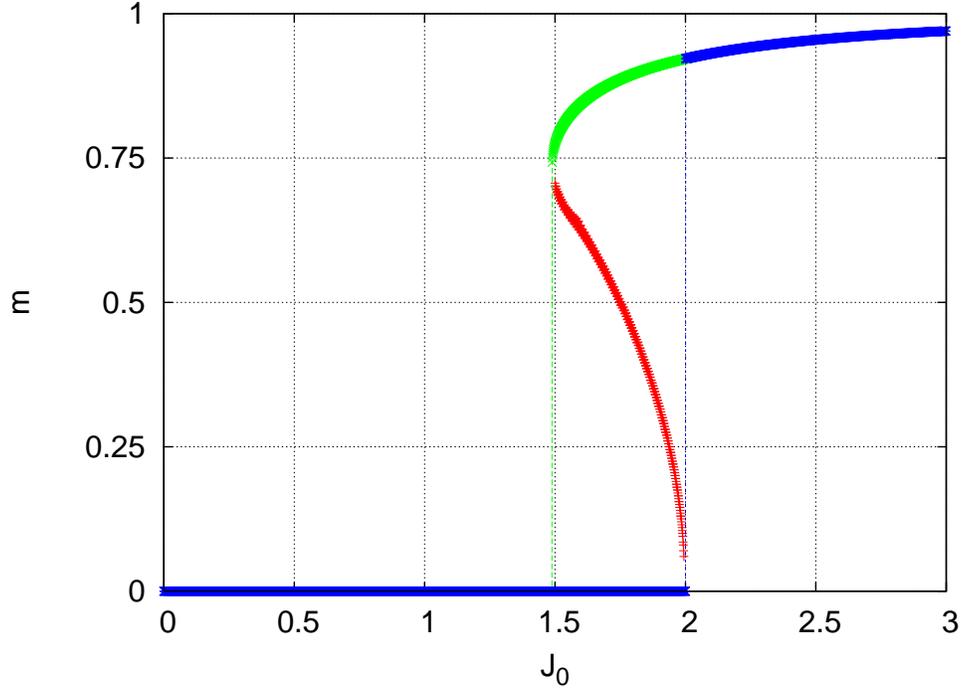}

\caption{Fixed points of the random field XY model in zero 
applied field and different values of exchange interaction 
$J_0$ in the mean field approximation. The blue curve 
corresponds to increasing $J_0$; at each value of $J_0$, the 
fixed point configuration of the previous lower value of 
$J_0$ is used as an input into the equations of dynamics. For 
increasing $J_0$, the magnetization of the fixed point is 
zero in the range $0 \le J_0 \le 2$. At $J_0=2$, it jumps to 
$m \approx .92$, and follows the blue curve as $J_0$ is 
further increased. The return path in decreasing $J_0$ is 
identical with the blue curve up to $J_0\le2$, but it 
continues along the green curve up to $J_0 \approx 1.49$ at 
which point it jumps from $m \approx .74$ to $m=0$ and 
remains zero for $0 \le J_0 \le 1.49 $. The red curve shows a 
set of unstable fixed points.} \label{fig2} \end{figure}

\begin{figure}[p] 
\includegraphics[width=.75\textwidth,angle=0]{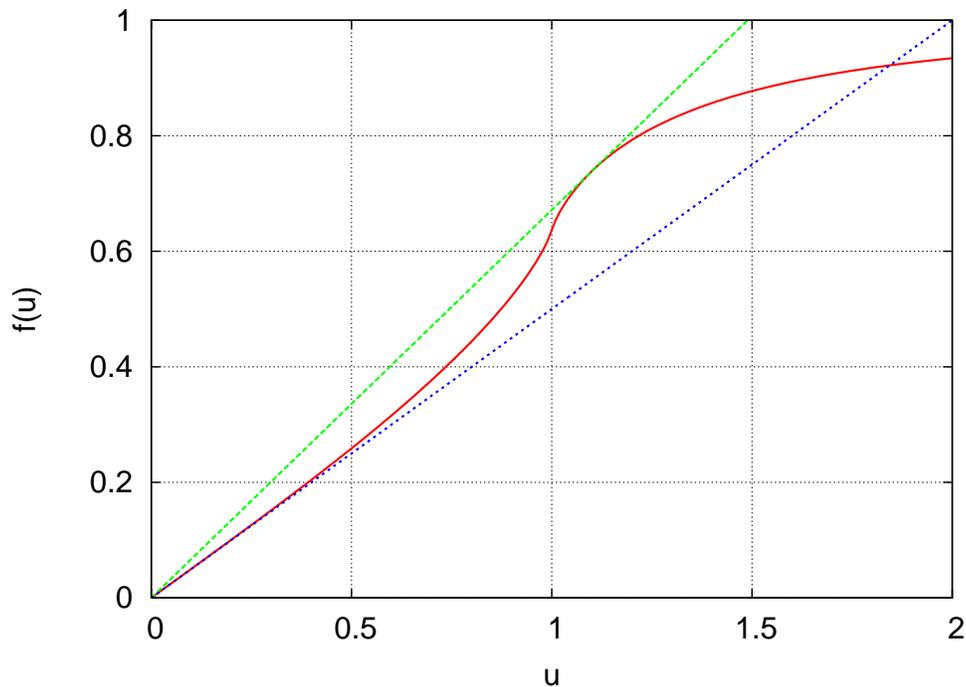}

\caption{ Graph of $f(u)$ vs $u$; and $u/J_0$ vs $u$ for 
$J_0=1.489$, and $J_0=2$ respectively. The figure shows why 
the mean field dynamics of the random field XY model has 
multiple fixed points in the range $1.489 \le J_0 \le 2$ as 
seen in figure 1.} \label{fig3} \end{figure}

\begin{figure}[p] 
\includegraphics[width=.75\textwidth,angle=0]{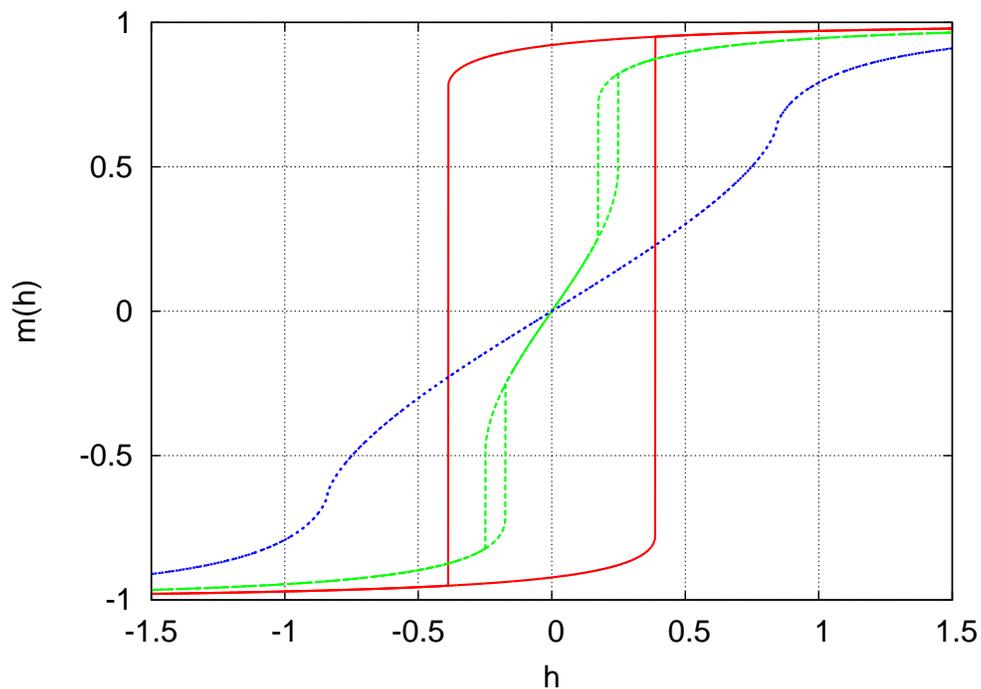}

\caption{ Hysteresis in the random field XY model in the mean 
field approximation. The figure shows magnetization in the 
system as the applied field $h$ is cycled along the $x$-axis 
for three representative values of $J_0$; $J_0=2$ (red), 
$J_0=1.25$ (green), and $J_0=.25$ (blue). The random field 
has a fixed magnitude equal to unity. The hysteresis loop for 
$J_0=2$ has a familiar shape but $J_0=1.25$ shows rather 
unusual hysteresis in two small windows of applied field 
situated at $h \approx -.2$, and $h \approx .2$ respectively 
but no hysteresis outside these windows. In particular, there 
is no hysteresis at or near zero applied field. For 
$J_0=.25$, there is no discernible hysteresis in any region 
of the applied field on the scale of the above figure.} 
\label{fig4} \end{figure}

\begin{figure}[p] 
\includegraphics[width=.75\textwidth,angle=0]{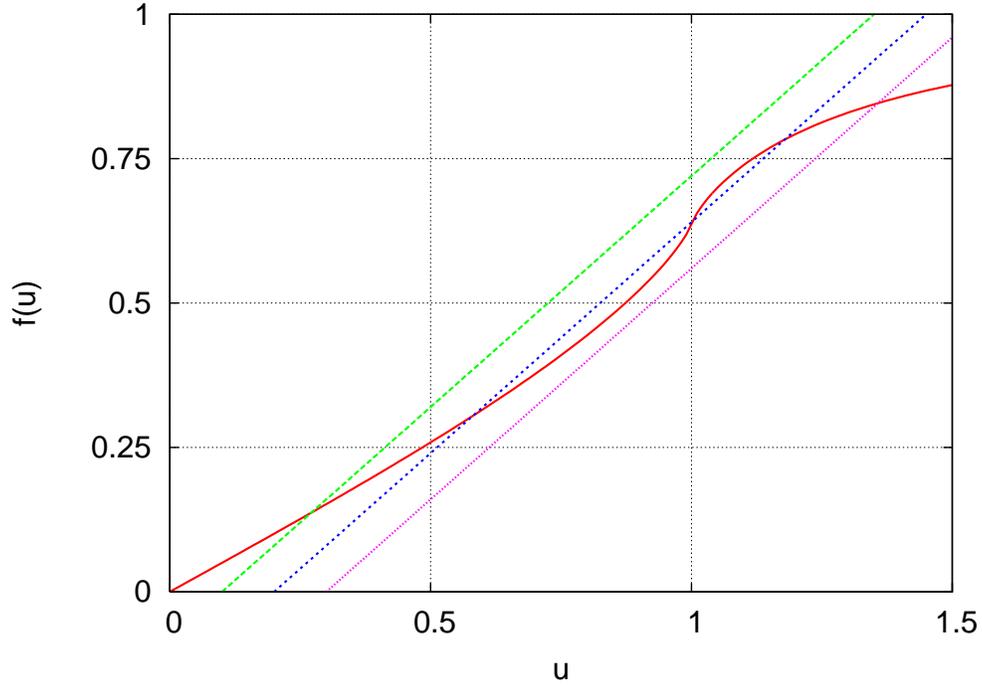}

\caption{Three straight lines $(u-h)/J_0$ for $J_0=1.25$ and 
$h=.1$, $h=.2$, $h=.3$ respectively. The lines are 
superimposed on a graph of $f(u)$ vs $u$ for $0 \le u \le 2$. 
Lines corresponding to $h=.1$ and $h=.3$ cut $f(u)$ at a 
single point but the line corresponding to $h=.2$ cuts $f(u)$ 
at three points. This accounts for two small hysteresis loops 
seen in figure (4) at $h=\pm .2$ and $J=1.25$.} \label{fig5} 
\end{figure}

\begin{figure}[p] 
\includegraphics[width=.75\textwidth,angle=0]{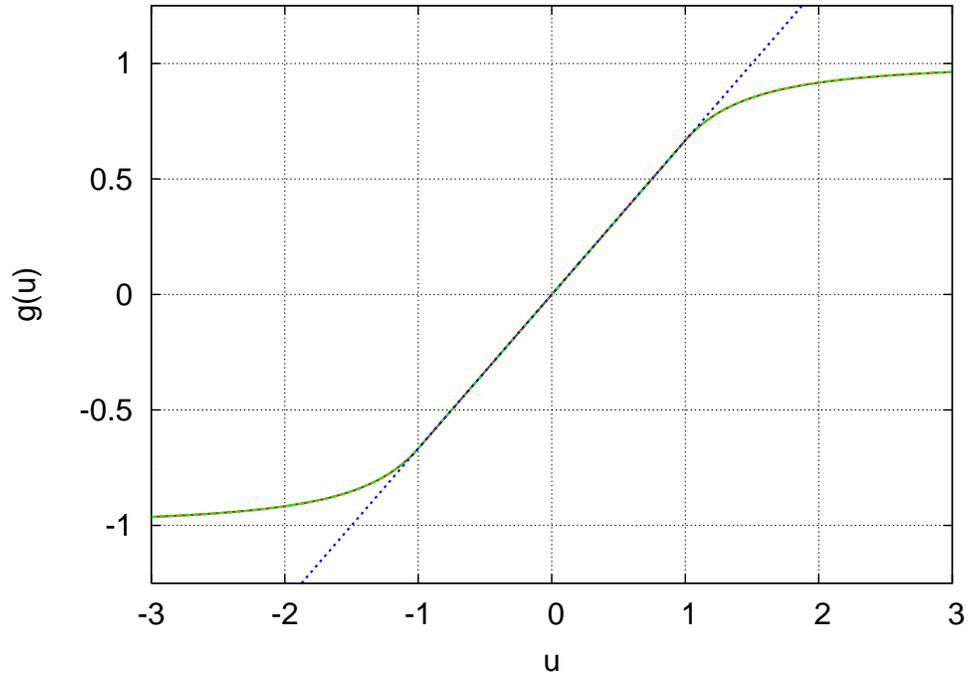} 
\caption{Graph of $g(u)$ vs $u$ for $-3 \le u \le 3$ 
superimposed on a line $u/J_0$ for $J_0=3/2$: $g(u)$ 
coincides with the line $2u/3$ in the range $-1 \le u \le 1$. 
If $J_0 < 3/2$, the line $u/J_0$ cuts $g(u)$ only at $u=0$. 
If $J_0 > 3/2$ the line $u/J_0$ cuts $g(u)$ at three points 
including the point $u=0$.} \label{fig6} \end{figure}

\begin{figure}[p] 
\includegraphics[width=.75\textwidth,angle=0]{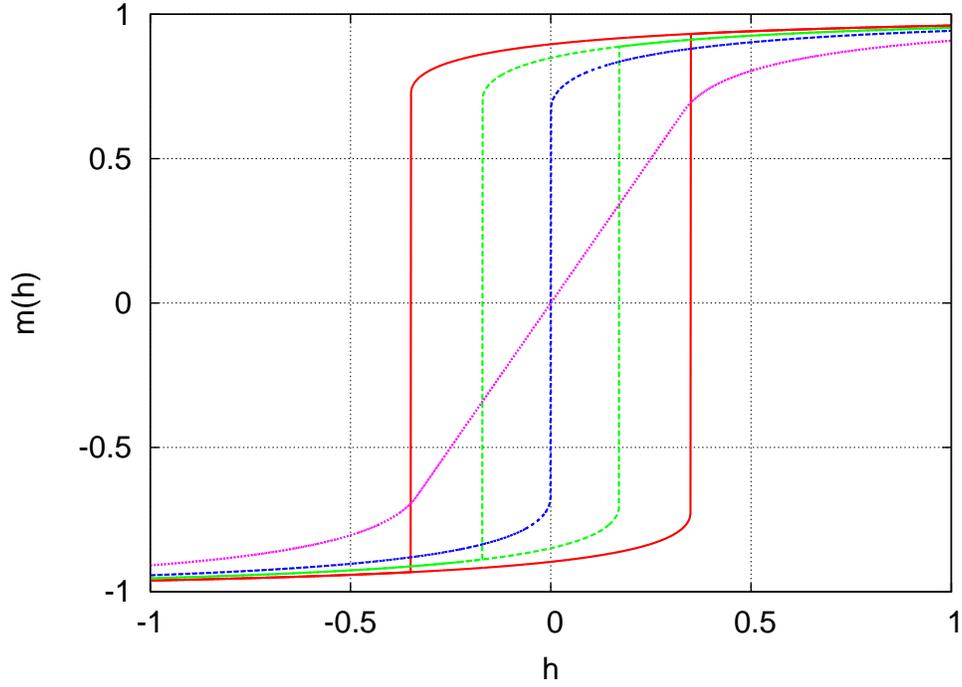}

\caption{ Hysteresis loops for the random field Heisenberg 
model for $J_0=2$ (red) and $J_0=1.75$ (green). There is no 
hysteresis if $J_0 \le 1.5$. Magnetization curves are shown 
for $J_0=1.5$ (blue) and $J_0=1$ (pink); magnetization may 
show a discontinuity at $h=0$ if $J_0=1.5$. }\label{fig7} 
\end{figure}

\begin{figure}[p] 
\includegraphics[width=.75\textwidth,angle=0]{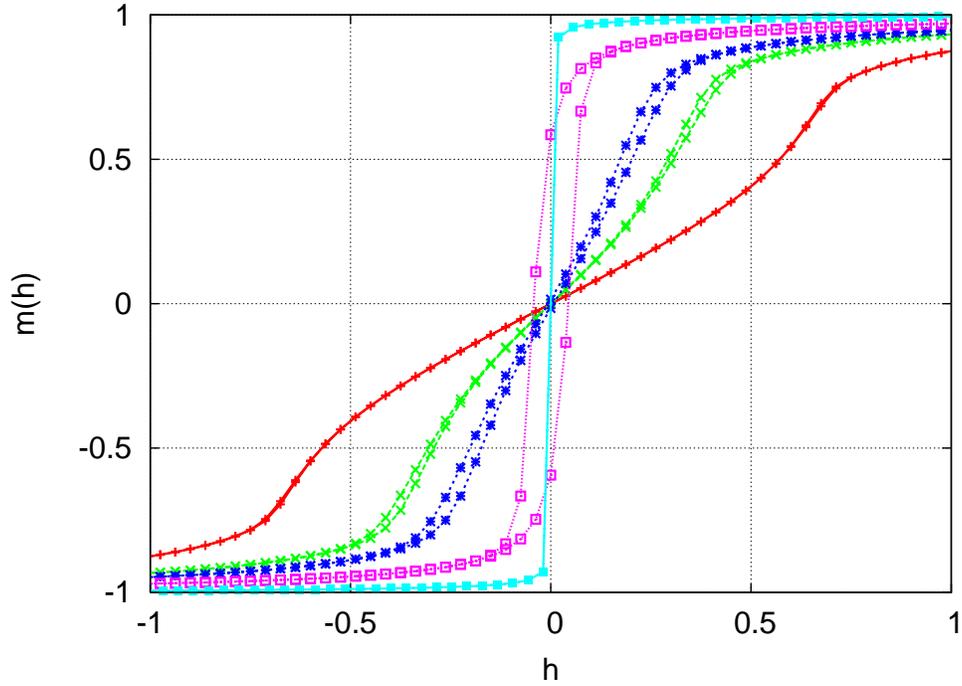} 
\caption{ Magnetization in the random field XY model on a 
$100^3$ simple cubic lattice under a cyclic field for 
different values of $J$: $J=.1$ (orange), $.2$ (green), $.25$ 
(blue), $.4$ (pink), and $1$ (light blue).} \label{fig8} 
\end{figure}

\begin{figure}[p] 
\includegraphics[width=.75\textwidth,angle=0]{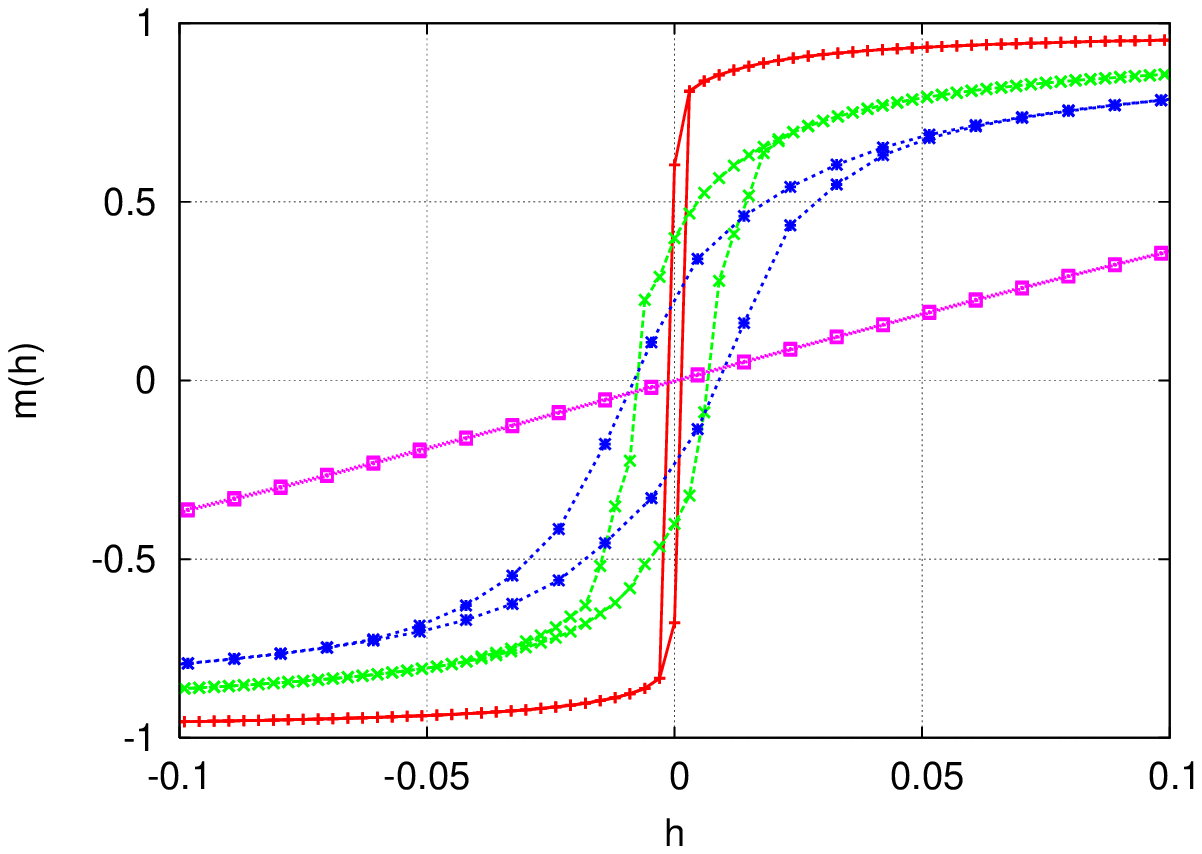} 
\caption{ Magnetization in the random field Heisenberg model 
on a $50^3$ simple cubic lattice under a cyclic field for 
different values of $J$: $J=.25$ (pink), $.4$ (blue), $.5$ 
(green), and $1$ (red).} \label{fig9} \end{figure}


\begin{thebibliography}{99}

\bibitem{imry-ma} Y Imry and S Ma Phys Rev Lett 35, 1399 
(1975); for a more recent review of random field phenomena 
see e.g. T Nattermann in
~\cite{young}.

\bibitem{young} {\em{Spinglasses and Random Fields}}, ed. A P 
Young, World Scientific, Singapore (1997).

\bibitem{biljakovic} K Biljakovic, in {\em{Phase Transitions 
and Relaxation in Systems with Competing Energy Scales}}, 
NATO Advanced Study Institute, Geilo, Norway (1993), eds. T 
Riste and D Sherrington, Kluwer Academic Publishers, 
Dordrecht (1993).

\bibitem{strogatz1} S H Strogatz, C M Marcus, and R M 
Westervelt, Phys Rev Lett 61, 2380 (1988).

\bibitem{gruner} G Gruner, Rev Mod Phys 60, 1129 (1988).

\bibitem{fisher} M P A Fisher, D S Fisher, and D A Huse, Phys 
Rev B 43, 130 (1990); D A Huse, M P A Fisher, and D S Fisher, 
Nature 358, 553 (1992).

\bibitem{giamarchi} T Giamarchi and S Bhattacharya, in 
{\em{High Magnetic Fields: Application to condensed matter 
physics and spectroscopy,}} eds. C Berthier, Springer-Verlag, 
2002.

\bibitem{ianna} G S Iannacchione et al, Phys Rev Lett 71, 
2595(1993); R L Leheny et al, Phys Rev E 67, 011708(2003).

\bibitem{maritan} A Maritan, M Cieplak, T Bellini, and R 
Banavar, Phys Rev Lett 72, 4113 (1994).

\bibitem{crawford} {\em{Liquid Crystals in Complex 
Geometries}}, eds. G P Crawford and S Zumer(Taylor \& 
Francis, London, 1996).

\bibitem{buscaglia} M Buscaglia et al, Phys Rev E 74, 011706 
(2006).

\bibitem{dieny} B Dieny and B Barbara, Phys Rev B 41, 11 549 
(1990); R Ribas, B Dieny, B Barbara, and A Labrata, J Phys: 
Condens Matter 7, 3301 (1994).

\bibitem{silveira1} R da Silveira and M Kardar, Phys Rev E 
59, 1355 (1999).


\bibitem{silveira2} R A da Silveira and S Zapperi, Phys Rev B 
69, 212404 (2004).

\bibitem{pierce1} M S Pierce et al, Phys Rev Lett 94, 017202 
(2005).

\bibitem{pierce2} M S Pierce et al, Phys Rev B 75, 144406 
(2007).

\bibitem{jagla} E A Jagla, Phys Rev B 72, 094406 (2005).

\bibitem{sethna1}J P Sethna, K A Dahmen, S Kartha, J A 
Krumhansl, B W Roberts, and J D Shore, Phys Rev Lett 70, 3347 
(1993); K Dahmen and J P Sethna, Phys Rev Lett 71, 
3222(1993); O Perkovic, K Dahmen, and J P Sethna, Phys Rev 
Lett 75, 4528 (1995).

\bibitem{sethna2} J P Sethna, K A Dahmen, and O. Percovic in 
{\em{The Science of Hysteresis}}, edited by G Bertotti and I 
Mayergoyz, Academic Press, Amsterdam (2006), and references 
therein.

\bibitem{dhar} D Dhar, P Shukla, and J P Sethna, J Phys A30, 
5259 (1997).

\bibitem{goicoechea} J Goicoechea and J Ortin, J Phys IV 
France 05, C2-71 (1995).

\bibitem{mirollo} R E Mirollo and S H Strogatz, SIAM J Appl 
Math 50, 108 (1990).

\bibitem{shukla} See for example, P Shukla and M S Green, 
Phys Rev Lett 34, 436(1975).

\bibitem{parisi} G Parisi and N Sourlas, Phys Rev Lett 43, 
744 (1979); also see G Parisi in {\em{Recent Advances in 
Field Theory and Statistical Mechanics}}, Proceedings of the 
Les Houches Summer School, Session XXXIX (North-Holland, 
Amsterdam, 1982), and references therein.

\bibitem{hartmann} A K Hartmann and U Novak, Eur Phys J B7, 
105 (1999).

\bibitem{bennett} L H Bennett and E D Torre, J Appld Phys 97, 
10E502 (2005)

\bibitem{cardone} D Cardone, M Dolce, and G Gesualdi, Bull 
Earthquake Eng 7, 801 (2009).

\bibitem{marchetti} M Cristina Marchetti, 
arXiv:cond-mat/0503660; M Cristina Marchetti, 
arXiv:cond-mat/0503639.

\bibitem{saunders} K Saunders, J M Schwarz, M C Marchetti, 
and A A Middleton, Phys Rv B 70, 24205 (2004).

\end{thebibliography}
\end{document}